\documentclass[aps,amsmath,amssymb,prl,reprint,longbibliography]{revtex4-2}
\usepackage{graphicx}

\usepackage{siunitx}

\begin{document}
\title{Paramagnetism in spherically confined charged active matter}
\author{Bal\'azs N\'emeth}
\email{bn273@cam.ac.uk}

\author{Ronojoy Adhikari}
\email{ra413@cam.ac.uk}

\affiliation{Department of Applied Mathematics and Theoretical Physics, Centre
for Mathematical Sciences, University of Cambridge, Wilberforce Road,
Cambridge CB3 0WA, England, United Kingdom}
\begin{abstract}
The celebrated theorem of Bohr and van Leeuwen guarantees that a classical
charged system cannot have a magnetization in thermal equilibrium.
Quantum mechanically, however, a \emph{diamagnetic} response is obtained.
In contrast, we show here that a classical charged active system,
consisting of a motile particle confined to the surface of a sphere,
has a nonzero magnetization and a \emph{paramagnetic} response. We
numerically sample Langevin trajectories of this system and compare
with limiting analytical solutions of the Fokker-Planck equation,
at small and large temperatures, to find excellent agreement in the
magnetic response. Our Letter suggests experimental routes to controlling
and extracting work from charged active matter.
\end{abstract}
\maketitle
\emph{Introduction. }The understanding that the net magnetization in a classical system
of charges in equilibrium must be zero can be traced to the work of
Bohr and van Leeuwen (BvL), who established this result by a consistent
application of classical electrodynamics and statistical mechanics
\cite{bohr_studies_1972,van_leeuwen_problemes_1921}. Peierls referred
to this as a ``surprise'' in theoretical physics, as it required
delicate cancellations of boundary currents only revealed through
a dynamical analysis \cite{peierls_surprises_1979}. Landau showed,
however, that by quantizing the motion of the charge a \emph{diamagnetic}
response could be obtained \cite{landau_diamagnetismus_1930}.

While the theorem of BvL rules out the possibility
of magnetization in a classical system in equilibrium, it is silent
on its magnetic response out of equilibrium. To drive a system out
of equilibrium requires a flux of energy that can be supplied externally
(as in a driven system) or internally (as in an active system). For
a particle system, both drive and activity can be encoded through
forces that, typically, lead to motility. A motile particle carrying
a charge will produce a spontaneous electric current. It is of both
theoretical and practical interest to ask if such microscopic active
currents can add coherently at finite temperatures to produce a macroscopic
magnetization.

To answer this question theoretically, it is desirable to focus on
a system that is compact and without boundary. Compactness ensures
that a nonequilibrium stationary state (NESS) can be reached, while
the absence of a boundary removes the need to study the ``surprising''
effects of boundary currents. The obvious geometry satisfying these
desiderata is the surface of a sphere. It has the additional advantage
of being easily realizable in experiment, for instance by confining
a charged motile particle to the surface of a liquid drop \cite{fei_magneto-capillary_2018}.
The facility of this geometry for theoretical analysis has already
been established \cite{kumar_classical_2009,pradhan_nonexistence_2010}.

In this Letter we consider the Langevin dynamics of a charged particle
confined to a spherical interface in a uniform magnetic field and
driven out of equilibrium by an active force that is constant in its
body frame. This requires both position and orientation for the description
of motion. We include, correspondingly, translational and rotational
inertia of the particle. The velocity dependence of the Lorentz force
requires us to retain the translational velocity as a dynamical variable
\cite{chun_emergence_2018,vuijk_anomalous_2019}. It is symmetrical
to also retain the angular velocity as a dynamical variable. This
allows us to formulate the dynamics geometrically in the phase space
$G\times\mathfrak{g}$, where $G$ is the Lie group of proper rotations
and $\mathfrak{g}$ is its Lie algebra. This also enables noise to
be added rigorously in the presence of constraints, avoids the Ito-Stratonovich
dilemma, and facilitates the accurate geometric numerical sampling
of Langevin trajectories. 

We confirm analytically that the magnetization vanishes at zero activity
for all temperatures, as expected from the equilibrium considerations
of BvL. In the presence of activity, the system reaches a NESS. From
the numerical sampling of trajectories, we find a nonzero net magnetization
in the direction of the applied field. This is confirmed by perturbative
analytical calculations at low and high temperatures. Strikingly,
the induced magnetization is \emph{paramagnetic}, in contrast to the
\emph{diamagnetic} response obtained quantum mechanically in the absence
of activity. The magnetization is produced by stochastically stable
Amperian loops of a given handedness that concentrate near the equator
at low temperatures. We now describe our results in detail.
\begin{figure}
  \includegraphics[width=\columnwidth]{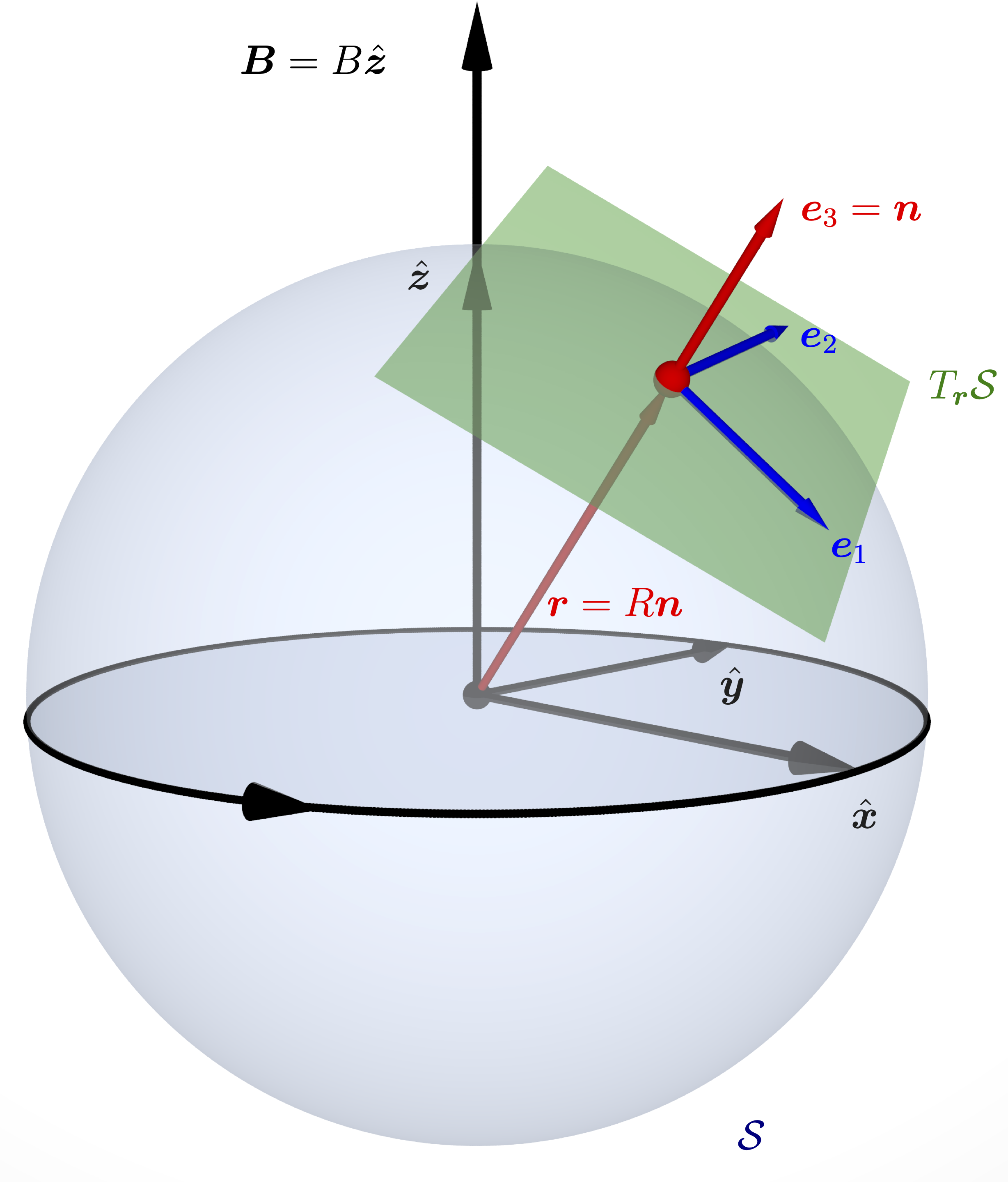}
\caption{A charged active particle at position $\boldsymbol{r}$ is confined
to a sphere $\mathcal{S}$ of radius $R$ in a uniform magnetic field
$\boldsymbol{B}=B\hat{\boldsymbol{z}}$. An orthonormal body frame
$(\boldsymbol{e}_{1},\boldsymbol{e}_{2},\boldsymbol{e}_{3})$ is rigidly
attached to the particle, with $\boldsymbol{e}_{3}$ normal to the
sphere. The velocity of the particle is in the plane $T_{\boldsymbol{r}}S$
tangent to $S$ at the point $\boldsymbol{r}$ and its angular velocity
is along $\boldsymbol{e}_{3}$. The configuration space is the Lie
group $G=SO(3)$ and the phase space is $G\times\mathfrak{g}$, where
$\mathfrak{g}$ is the Lie algebra of $G$. The solid line shows a
right-handed equatorial Amperian current loop.}
\end{figure}
\begin{figure*}
\includegraphics[width=0.25\textwidth]{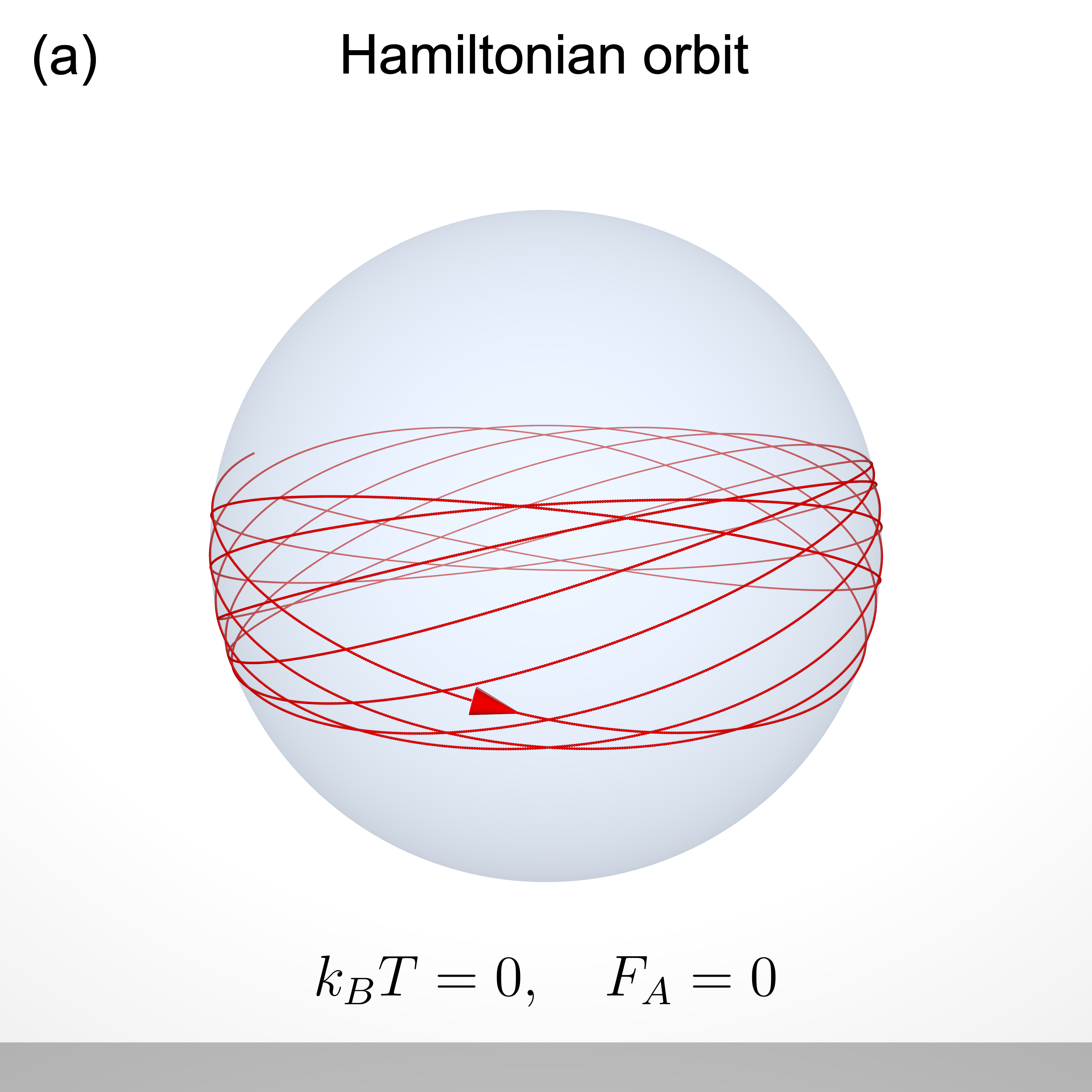}\includegraphics[width=0.25\textwidth]{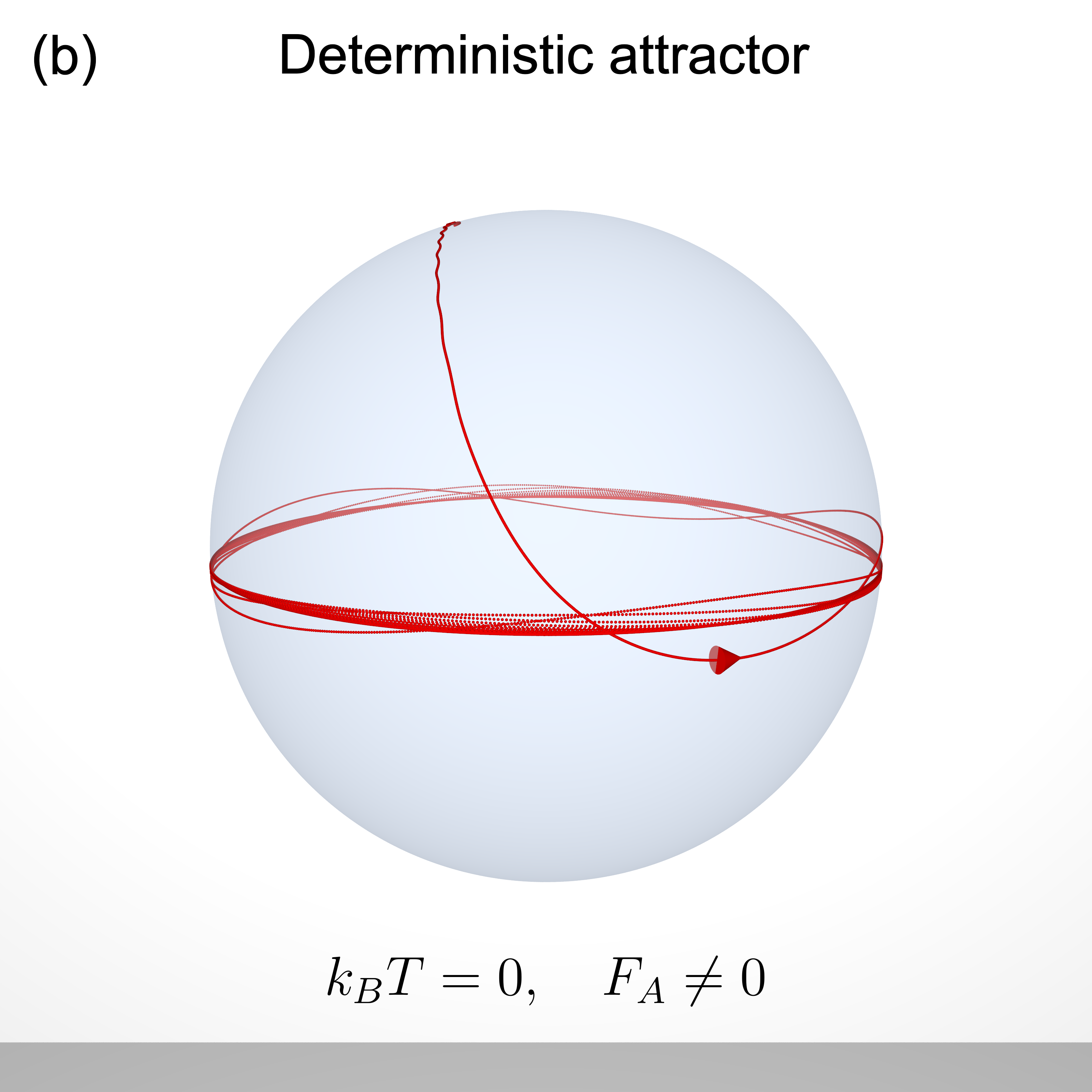}\includegraphics[width=0.25\textwidth]{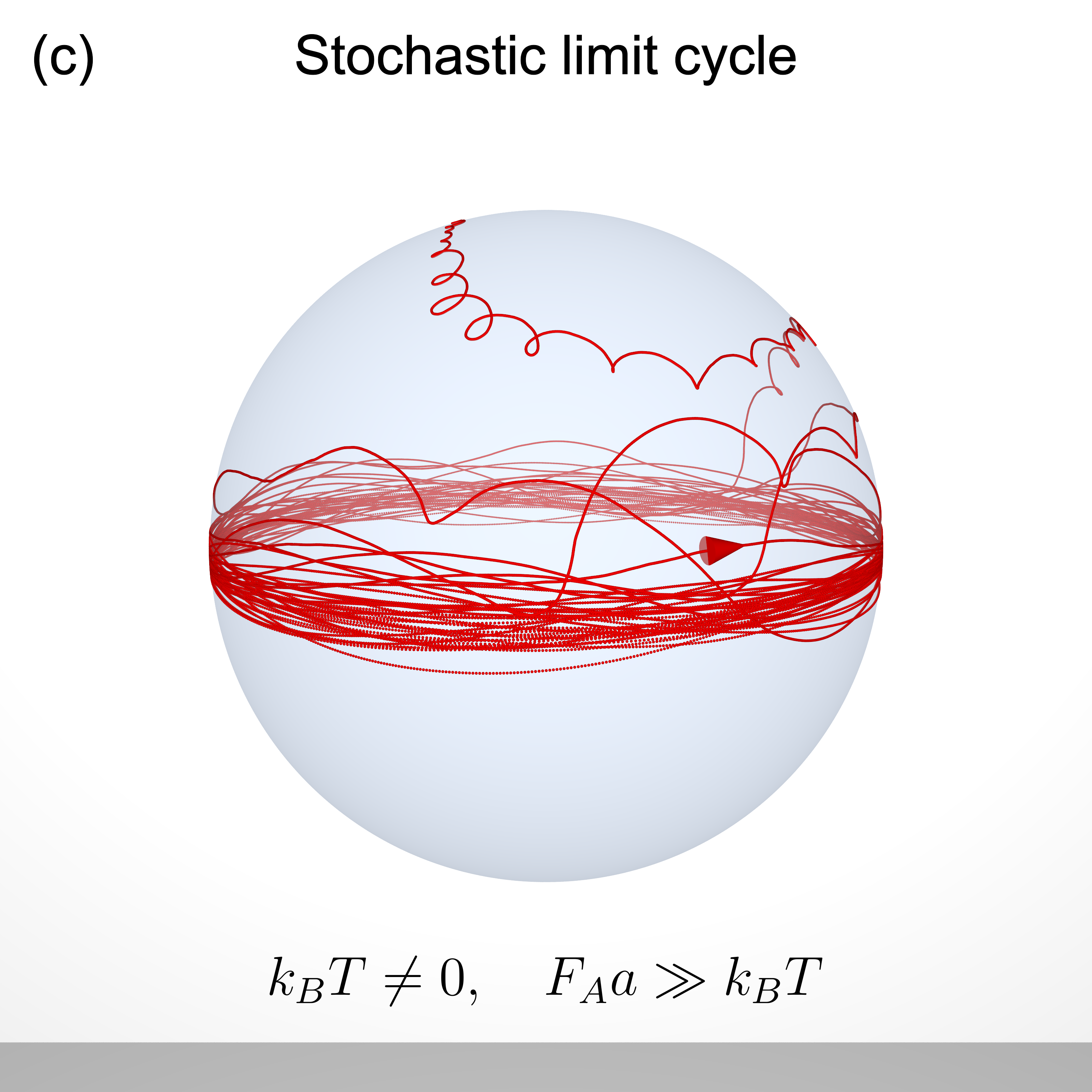}\includegraphics[width=0.25\textwidth]{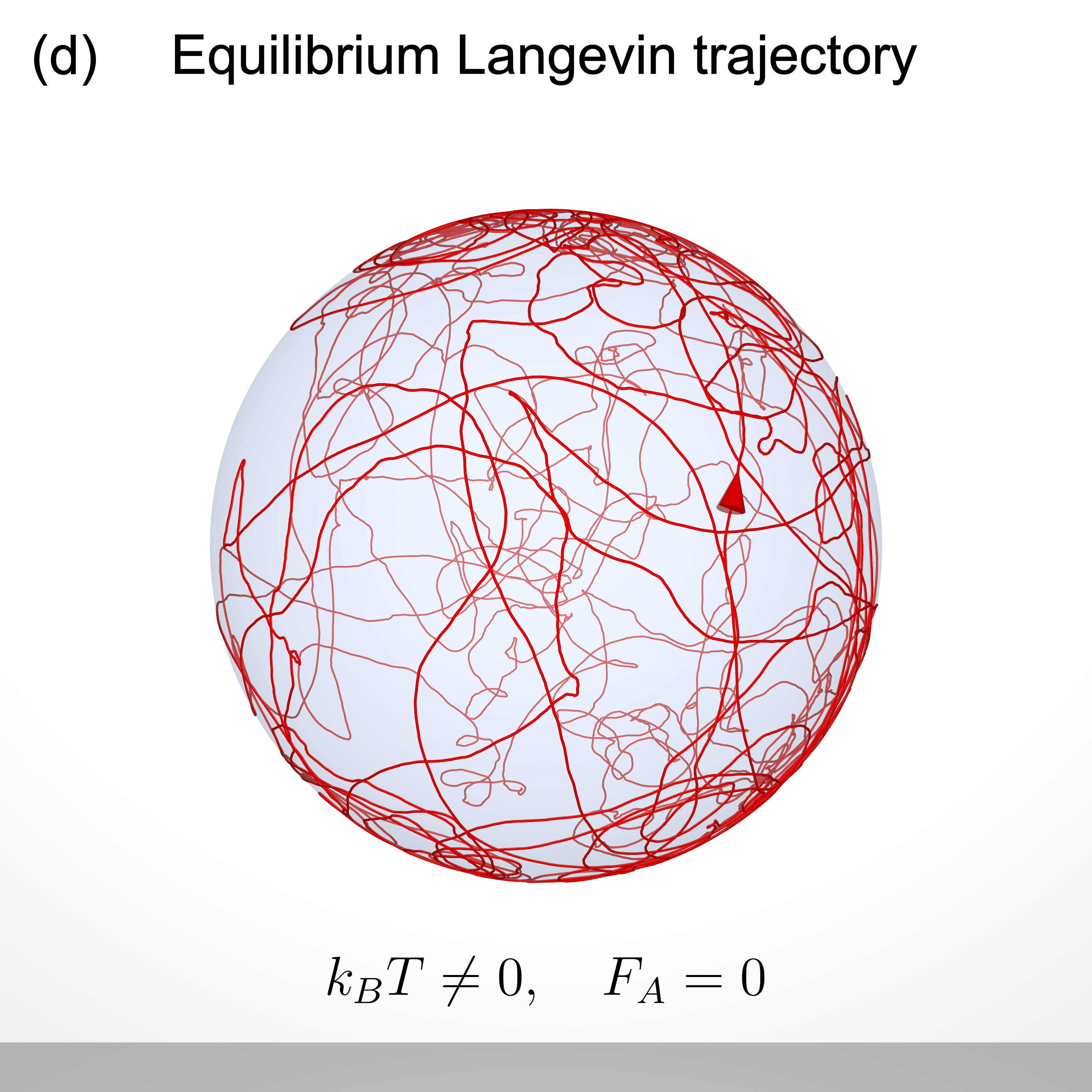}
\caption{Typical trajectories of deterministic ($k_{B}T=0)$
and stochastic ($k_{B}T\protect\neq0)$ dynamics for magnetic field
$B$ and activity $F_{A}$. (a) Hamiltonian orbits for $k_{B}T=F_{A}=0$ and zero friction $\gamma_T=\gamma_R=0$.
(b) Deterministic attractor for $k_{B}T=0$. (c) Stochastic limit
cycle at low temperatures with $F_{A}a\gg k_BT$. The deterministic
attractor is stochastically stable at low temperatures and trajectories
in (c) contribute to a nonzero paramagnetic response. (d) Equilibrium
Langevin trajectories $k_BT\neq 0,F_{A}=0$ on the sphere. See movie
in Supplemental Material \cite{supp} for animations of the dynamics.}
\end{figure*}

\emph{Kinematics. }We model the active particle as a rigid body
of size $a$ the center of mass of which is located at $\boldsymbol{r}$
and the principal axes of inertia of which are the orthonormal triad of frame
vectors $\boldsymbol{e}_{i}$ with $i=1,2,3$. The particle is constrained
to move on the surface of a sphere $\mathcal{S}$ of radius $R$ with
its $\boldsymbol{e}_{3}$ axis pointing along the outward normal $\boldsymbol{n}$
of the sphere. Since $\boldsymbol{e}_{3}$ is normal to the sphere,
the vectors $\boldsymbol{e}_{1}$ and $\boldsymbol{e}_{2}$ span the
tangent plane $T_{\boldsymbol{r}}\mathcal{S}$ and provide a moving
frame for resolving vectors (see Fig. (1)). The velocity $\boldsymbol{v}$
and the angular velocity $\boldsymbol{\Omega}$, defined by $d\boldsymbol{r}=\boldsymbol{v}dt$
and $d\boldsymbol{e}_{i}=\left(\boldsymbol{\Omega}dt\right)\times\boldsymbol{e}_{i}$,
are resolved in the moving frame as $\boldsymbol{v}=v_{i}\boldsymbol{e}_{i}$
and $\boldsymbol{\Omega}=\Omega_{i}\boldsymbol{e}_{i}$, where repeated
indices are summed. The motional constraints are implemented by setting
$v_{3}=0$ (no normal translation) and $\boldsymbol{r}\equiv R\boldsymbol{n}=R\boldsymbol{e}_{3}$
(no independent rotation about an axis in the tangent plane). The
particle configuration, then, is fully contained in the three frame
vectors. From the rotational constraint it follows that $d\boldsymbol{r}=Rd\boldsymbol{e}_{3}=R(\boldsymbol{\Omega}dt)\times\boldsymbol{e}_{3}$
and so $\boldsymbol{v}=R\boldsymbol{\Omega}\times\boldsymbol{e}_{3}$,
showing that the tangential components of the angular velocity and
the translational velocity are mutually dependent. This leaves three
independent kinematic degrees of freedom, which we take to be the
two components of translational velocity $v_{1}$ and $v_{2}$ and
the single component of angular velocity $\omega\equiv\Omega_{3}$,
all expressed in the body frame. Since the motion of the frame is
an orthogonal transformation, the configuration space can be identified
with the special orthogonal group $G=SO(3)$. The infinitesimal change
in the frame $\boldsymbol{\Omega}dt$ is then naturally identified
with the Lie algebra $\mathfrak{g}=\mathfrak{so}(3)$ of infinitesimal
rotations. The dynamical phase space is $G\times\mathfrak{g}$. The
\emph{constrained }motion of the particle can thus be described by the
\emph{unconstrained} motion of its body frame. This offers significant
advantages in formulating the equations of motion as we show below
\cite{poincare_sur_1901}.

\emph{Deterministic dynamics. }We immerse the system above in a
uniform magnetic field $\boldsymbol{B}=B\hat{\boldsymbol{z}}$ with
a corresponding vector potential $\boldsymbol{A}$, such that $\boldsymbol{B}=\nabla\times\boldsymbol{A}$.
The particle has mass $m$ and moment of inertia $J$ along the $\boldsymbol{e}_{3}$
axis. The deterministic dynamics is most easily formulated in terms
of the Lagrangian $L=\frac{1}{2}m\boldsymbol{v}\cdot\boldsymbol{v}+\frac{1}{2}J\omega^{2}+q\boldsymbol{A}\cdot\boldsymbol{v}$, which includes translational and rotational kinetic energies and the
coupling to the magnetic field. We vary this Lagrangian maintaining
the orthogonality of the frame vectors and respecting the velocity
constraints to obtain the conservative dynamics \cite{supp,marsden_introduction_2003}.
To this we add an active force $\boldsymbol{F}_{A}=F_{A}\boldsymbol{e}_{1}$
that is constant in the body frame and acts along the first principal
axis, frictional forces proportional to the velocities and a frictional
torque proportional to the angular velocity. The resulting equations
are

\begin{alignat}{1}
m\dot{v}_{1} & =+\left(\Delta m\omega+m\omega_{c}\frac{z}{R}\right)v_{2}-\gamma_{T}v_{1}+F_{A},\label{eq:euler1}\\
m\dot{v_{2}} & =-\left(\Delta m\omega+m\omega_{c}\frac{z}{R}\right)v_{1}-\gamma_{T}v_{2},\label{eq:euler2}\\
J\dot{\omega} & =-\gamma_{R}\omega,\quad\omega_{c}=\frac{qB}{m},\label{eq:euler3}\\
\dot{\boldsymbol{e}_{i}} & =\boldsymbol{\Omega}\times\boldsymbol{e}_{i},\quad\boldsymbol{e}_{3}=\frac{\boldsymbol{r}}{R},\label{eq:kinematics}
\end{alignat}
where $\Delta m=m\left(1-\frac{J}{mR^{2}}\right)$ is an effective
mass, $\omega_{c}$ is the cyclotron frequency, and $\gamma_{T}$
($\gamma_{R}$) is the translational (rotational) friction coefficient.
Since $J\sim ma^{2}$, we can neglect the correction to the mass for
a small particle on a large sphere $(a\ll R)$ and set $\Delta m=m.$
Our results do not change qualitatively by making this approximation.
Then, the conservative part of the above dynamics rotates velocities
at a rate $\varpi=\omega+\omega_{c}z/R$ and conserves the angular
velocity $\omega$. These Hamiltonian orbits, shown in Fig. 2(a), are
similar to those of a symmetric top but show a precession about the
axis of the external magnetic field. With friction and activity, the
Hamiltonian orbits are attracted to a limit cycle at the equator with
$v_{1}=F_{A}/\gamma_{T}\equiv u$, $v_{2}=0$ and $\omega=0$, as
shown in Fig. 2(b). The equatorial limit cycles are reminiscent of
traveling bands of active particles around geodesics reported in earlier
studies of many interacting active agents on spheres \cite{sknepnek_active_2015,henkes_dynamical_2018}.
There, a stable equatorial orbit is chosen through a spontaneous and
collective many-particle breaking of symmetry, while here the external
magnetic field breaks symmetry and selects the stable equatorial orbit
even at the one-particle level. 

The limit cycle results in an Amperian loop carrying a current $qu\big/2\pi R$
and enclosing a planar area $\pi R^{2}$ which produces a magnetic
moment $\mu_{a}=quR/2$ in the $z$ direction. The moment is parallel
to the applied field and hence the response is paramagnetic. Thus,
the interplay of an applied magnetic field, activity, friction, and
confinement yields an activity-induced realization of a classical
magnetic spin. How does this effective spin behave at finite temperatures?

\emph{Stochastic dynamics. }To answer this question we promote our
deterministic equations of motion to Langevin equations by adding
noise that obeys the fluctuation-dissipation relation (FDR). This
is in contrast to \cite{muhsin_orbital_2021}, where the authors
find a nonzero magnetization in a NESS obtained by violating the
FDR. To make contact with statistical mechanics, we introduce the
probability measure in phase space $p\left(\boldsymbol{e}_{i},\boldsymbol{\Omega},t\right)dgd\boldsymbol{\Omega}$,
where $dg$ is the invariant volume element in $G$ and $d\boldsymbol{\Omega}=dv_{1}dv_{2}d\omega$
is the volume element in $\mathfrak{g}$. The time evolution of the
probability density $p\left(\boldsymbol{e}_{i},\boldsymbol{\Omega},t\right)$
is governed by the Fokker-Planck equation $\partial_{t}p=\mathcal{L}p$,
derived from the Langevin equations via standard methods \cite{gardiner_stochastic_2009}.
The resulting Fokker-Planck operator $\mathcal{L}$ is \cite{supp}
\begin{widetext}
\begin{equation}
\mathcal{L}=\frac{v_{2}}{R}L_{1}-\frac{v_{1}}{R}L_{2}-\omega L_{3}+\varpi\left(\boldsymbol{v}\times\frac{\partial}{\partial\boldsymbol{v}}\right)+\frac{\gamma_{T}}{m}\frac{\partial}{\partial\boldsymbol{v}}\boldsymbol{v}+\frac{\gamma_{R}}{J}\frac{\partial}{\partial\omega}\omega-\frac{F_{A}}{m}\frac{\partial}{\partial v_{1}}+\frac{k_BT\gamma_T}{m^2}\frac{\partial^{2}}{\partial\boldsymbol{v}^{2}}+\frac{k_BT\gamma_R}{J^2}\frac{\partial^{2}}{\partial\omega^{2}}.\label{eq:fp_op}
\end{equation}
\end{widetext}The $L_{i}$ above are generators of infinitesimal
rotations about the frame vectors $\boldsymbol{e}_{i}$. They are
left-invariant vector fields in $G$ and are related to angular momentum
operators in quantum mechanics. The shorthand notations $\boldsymbol{v}\times\frac{\partial}{\partial\boldsymbol{v}}=v_{1}\frac{\partial}{\partial v_{2}}-v_{2}\frac{\partial}{\partial v_{1}}$,
$\frac{\partial}{\partial\boldsymbol{v}}\boldsymbol{v}=\frac{\partial}{\partial v_{1}}v_{1}+\frac{\partial}{\partial v_{2}}v_{2}$, and $\frac{\partial^{2}}{\partial\boldsymbol{v}^{2}}=\frac{\partial^{2}}{\partial v_{1}^{2}}+\frac{\partial^{2}}{\partial v_{2}^{2}}$
are used for brevity. The use of the body frame and the phase space
$G\times\mathfrak{g}$ yields a Fokker-Planck operator that is invariant
under choice of local coordinates. The subtle issues of adding noise
to degrees of freedom constrained to remain on manifolds ($G$ is
an orthogonal matrix group) are obviated by adding noise to their
velocities (which are unconstrained in the matrix Lie algebra $\mathfrak{g})$.
This also removes room for Ito-Stratonovich ambiguities. Langevin
trajectories can be sampled by splitting the Fokker-Planck operator
into the conservative Hamiltonian drift and an Ornstein-Uhlenbeck
drift diffusion in the space of angular velocities. Each of these
can be sampled efficiently and accurately \cite{hairer_geometric_2006,chin_symplectic_2008}.
The splitting method converges with a weak order of 2 and respects
the spherical confinement of the position and orthogonality of the
body frame. The magnetic moment is calculated from the average $\boldsymbol{\mu}=\left(q/2\right)\left\langle \boldsymbol{r}\times\boldsymbol{v}\right\rangle _{SS}$
with respect to the steady state probability measure, either analytically
or from numerically sampled trajectories. By symmetry, $\boldsymbol{\mu}$
can only have a nonzero component in the $z$-direction. We, therefore,
confine attention to the rescaled magnetic moment $\mu\equiv\left(\boldsymbol{\mu}\cdot\hat{\boldsymbol{z}}\right)\big/\mu_{a}$
and study its variation with the temperature and magnetic field. 

\emph{Thermal equilibrium. }We assume that the dynamics relaxes to
a unique stationary state $p_{SS}\left(\boldsymbol{e}_{i},\boldsymbol{\Omega}\right)$
as $t\rightarrow\infty$ which satisfies $\mathcal{L}p_{SS}=0$. In
the presence of the magnetic field but in the absence of activity
it is easy to verify by substitution that the Gibbs distribution,
\[
p_{SS}^{eq}\left(\boldsymbol{e}_{i},\boldsymbol{\Omega}\right)=\sqrt{\frac{m^{2}J}{\left(2\pi k_{B}T\right)^{3}}}\exp\left(-\frac{m\boldsymbol{v}^{2}+J\omega^{2}}{2k_{B}T}\right),
\]
is the stationary solution. This shows that $\boldsymbol{r}$ and
$\boldsymbol{v}$ are independently distributed in equilibrium and
it immediately follows that $\boldsymbol{\mu}^{eq}=\left(q/2\right)\left(\left\langle \boldsymbol{r}\right\rangle ^{eq}\times\left\langle \boldsymbol{v}\right\rangle ^{eq}\right)=0$.
This verifies the BvL theorem and confirms, through an independent
analysis, that there is no magnetization in equilibrium in a spherical
geometry \cite{kumar_classical_2009,pradhan_nonexistence_2010}.
We now turn to dynamics with activity.

\emph{Low temperatures and strong magnetic fields. }Since solving the Fokker-Planck equation analytically is no longer possible in the
presence of activity, we first examine the low-temperature limit,
asking if the equatorial attractor is stable to noise. We use a van
Kampen expansion \cite{van_kampen_stochastic_2007} to obtain an
Ornstein-Uhlenbeck process for small perturbations about the attractor.
We find the attractor to be stable to small noise and consequently
the Amperian loops, now stochastic, add coherently to produce a nonzero
magnetic moment. In the limit of vanishing rotational inertia, we
obtain to leading order in $k_{B}T$ \cite{supp}
\begin{equation}
\mu\approx1-\frac{k_{B}T}{quRB}.\label{eq:gauss_approx}
\end{equation}
This magnetic response is identical to the low-temperature limit of
a classical spin of magnetic moment $\mu_{a}$ in a constant magnetic
field of magnitude $B$ in thermal equilibrium. The average magnetic
moment $\mu_{a}$ is reduced by an amount proportional to $k_{B}T$
due to Gaussian fluctuations of the spin about the direction of the
field. The numerical sampling of trajectories confirms this analytical
result. Typical low-temperature trajectories, shown in Fig. 2(c), remain
confined near the attractor with $\langle z\rangle=0$ and $\langle z^{2}\rangle=k_{B}TR\big/qBu$.
The appearance of $B$ in the denominator of the expression for $\mu$
and $\langle z^{2}\rangle$ shows that in order to have a substantial
magnetic moment and a stable equatorial orbit, we need both low temperatures
and strong fields. These ensure a strongly attracting equatorial limit
cycle. The average magnetic moment computed from the trajectories
agrees very well with the analytical approximations as shown in Fig. 3. From Eq. (\ref{eq:gauss_approx}) we see that thermal fluctuations
are significant for $k_{B}T\sim quRB$. 
\begin{figure}
\centering{}\includegraphics[width=1\columnwidth]{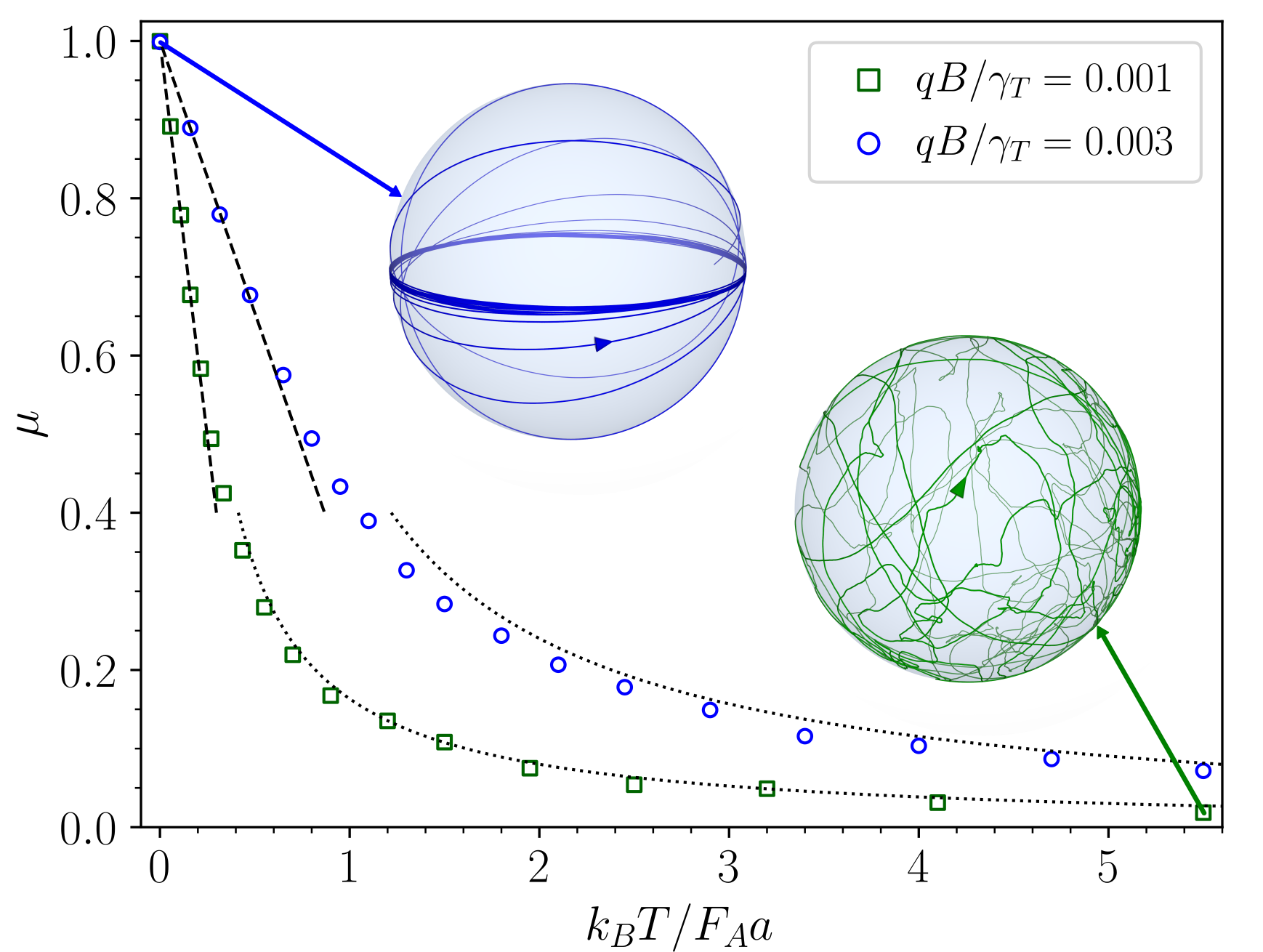}\caption{Magnetic moment $\mu$ as a function of temperature for fixed activity
and magnetic field. For low temperatures, the magnetic moment is a
linear function of temperature with intercept $1$, while for high
temperatures it decays as $1/k_{B}T$. The two limiting regimes are
separated by an intermediate region where $k_{B}T\sim quRB$. The
analytical predictions for the slope of the linear regime and the
rate of decay, shown as the dashed and dotted curves, respectively,
agree well with the numerical results. The insets corresponding to
data points with blue and green trajectories show low- and high-temperature
orbits, respectively. Error bars are bounded by the sizes of the markers. As the limit of zero rotational inertia is inaccessible
in our simulations, we have used the results of the perturbative calculations
at finite rotational inertia \cite{supp}.}
\end{figure}

\emph{High temperatures and weak magnetic fields. }Having established
a nonzero magnetization in the NESS at low temperatures and high fields,
we ask how robust this effect is to an increase in temperature or
a weakening of the field. This requires an involved calculation setting up the Brinkman hierarchy \cite{brinkman_brownian_1956} for the Fokker-Planck
operator (\ref{eq:fp_op}) and applying a closure. Again in the limit of vanishing rotational
inertia, we obtain to leading order in $1/k_{B}T$ and $B$ \cite{supp},
\begin{equation}
\mu\approx\frac{quRB}{3k_{B}T}.\label{eq:pert_magn_mom}
\end{equation}
The dependence of the magnetic moment on the magnetic field and temperature
is reminiscent of Curie's law in classical paramagnetism \cite{langevin_sur_1905,van_vleck_theory_1965}.

\emph{Irreversibility. }We now provide a trajectorial interpretation
of the nonzero magnetic moment. Assuming that the stochastic dynamics
is ergodic, we can write the steady state magnetic moment as a long-time
average of the path-dependent observable:\begin{subequations}
\begin{alignat}{1}
\mu & =\frac{\hat{\boldsymbol{z}}}{\mu_{a}}\cdot\left(\lim_{t\to\infty}\frac{q}{2t}\int_{0}^{t}\left(\boldsymbol{r}\times\boldsymbol{v}\right)dt\right)\label{eq:irrev1}\\
 & =\frac{1}{\mu_{a}}\left(\lim_{t\to\infty}\frac{q}{2t}\int_{\mathcal{C}_{t}}\left(\hat{\boldsymbol{z}}\times\boldsymbol{r}\right)\cdot d\boldsymbol{r}\right)\label{eq:irrev2}\\
 & \propto\lim_{t\to\infty}\frac{1}{t}\int_{\mathcal{C}_{t}}\boldsymbol{A}\cdot d\boldsymbol{r},\label{eq:irrev3}
\end{alignat}
\end{subequations}where in Eqs. (\ref{eq:irrev2}) and (\ref{eq:irrev3}),
the integral is along the stochastic trajectory $\mathcal{C}_{t}$
of duration $t$. We have made a gauge choice $\boldsymbol{A}(\boldsymbol{r})=B\left(\hat{\boldsymbol{z}}\times\boldsymbol{r}\right)\big/2$
but the result above is gauge independent. Equation (\ref{eq:irrev3})
shows that the steady state magnetic moment is proportional to the
long-time average of the line integral of the gauge potential along
the stochastic trajectory. This suggests that it is also a dynamical
measure of irreversibility in the system, where detailed balance is
broken by the nonconservative active force which is able to do work
on the particle when going around a closed loop in configuration space.
The instantaneous value of the integrand for trajectories in Fig. 2
is animated in \cite{supp}. The violation of detailed balance does
not necessarily imply a nonzero magnetic moment though, as the following
counterexample demonstrates. Replacing the uniform magnetic field
by that of a monopole at the center of the sphere produces a NESS
in which there is no breaking of symmetry and the position and the
velocity remain independently distributed. The magnetic moment thus
vanishes even in NESS. The subtleties of NESS in the presence of magnetic
fields have been noted in \cite{lee_nonequilibrium_2019,vuijk_lorentz_2020,abdoli_stationary_2020,abdoli_correlations_2020,matevosyan_nonequilibrium_2021,muhsin_inertial_2022,muhsin_active_2025}.

\emph{Conclusion. }We have shown that a charged active particle in
a confined geometry subject to linear friction follows a geodesic
that is selected by the magnetic field. This establishes an Amperian
loop pointing in the direction of the magnetic field and yields a
paramagnetic response at finite temperatures. This provides, both
qualitatively and quantitatively, a classical realization of spin.
A possible experimental realization of our system could be based on
charged nanomotors confined by capillary effects to spherical liquid-liquid
interfaces, for example the interfaces of emulsion droplets \cite{fei_magneto-capillary_2018}.
For micron-sized objects $a\approx\qty{1}{\micro\metre}$ with swimming
speed $u\approx\qty{100}{\micro\metre\per\second}$ \cite{wang_autonomous_2012}
at room temperature $T\approx\qty{300}{\kelvin}$, assuming Stokes drag
in a fluid with the viscosity of water, we find $F_{A}a\big/k_{B}T\approx 500$.
The relative variance of trajectories $\langle z^{2}/R^{2}\rangle=k_{B}T/quRB$
can be made small by increasing the magnetic field. We thus expect
the Amperian loops to be observable in room-temperature experiments.

The stability of active current loops can be harnessed to steer active
charged particles using a magnetic field \cite{davis_active_2024}.
Our system demonstrates an alternative way to control active particles using
magnetic fields via the Lorentz force compared to previous approaches
relying on magnetic torques and forces on permanent magnetic moments
\cite{tierno_controlled_2008,guillamat_control_2016,mandal_magnetic_2018,demirors_active_2018,bente_biohybrid_2018,matsunaga_controlling_2019}.
The present theory will apply if changes in the steering magnetic
field are slow compared to the relaxation time of the plane of the current
loop to a change in direction of the magnetic field. If the changes
are comparable, we can expect the transient between stationary states,
not studied here, to lead to memory effects \cite{thery_controlling_2024}
and a frequency-dependent paramagnetic response \cite{baiesi_fluctuations_2009}.
The latter suggests that nonequilibrium work can be extracted from
the system using suitable cyclic, and not necessarily adiabatic, time-dependent
magnetic fields \cite{fodor_active_2021,abdoli_tunable_2022,davis_active_2024}.
This further suggests interesting connections to stochastic thermodynamics
and fluctuation theorems, where electromechanical effects have been
noted \cite{jayannavar_charged_2007,saha_nonequilibrium_2008,pradhan_nonexistence_2010,jimenez-aquino_work-fluctuation_2011,seifert_stochastic_2012}.
The connections between our results and the rich phenomenology found
in chiral active matter \cite{lauga_swimming_2006,liebchen_chiral_2022}
remain to be explored. In conclusion, our work establishes the possibility
of paramagnetism in a classical system and reveals the surprising
effects that can arise through the interplay of electromagnetic, viscous, and active forces in soft matter.

\emph{Acknowledgments. }We thank Prof. M. E. Cates for insightful
comments and a careful reading of the Letter and Prof. S.
R. Ramaswamy for stimulating discussions. This research was supported
by Engineering and Physical Sciences Research Council Grant No. EP/W524141/1 (B.N.) and in part by NSF Grant No. PHY-2309135 to the Kavli
Institute for Theoretical Physics.

\end{document}